\documentclass{ifacconf}

\usepackage{natbib}        
\usepackage{amsmath,amssymb,amsfonts}
\usepackage{algorithmic}
\usepackage{graphicx}
\usepackage{textcomp}
\usepackage{xcolor}
\usepackage{mathtools}
\usepackage{epstopdf}
\usepackage{caption}
\usepackage{float}
\usepackage{subcaption}
\usepackage{dsfont}
\usepackage{lipsum,graphicx,subcaption}
\usepackage{ntheorem}

\makeatletter
\renewcommand*\env@matrix[1][*\c@MaxMatrixCols c]{%
  \hskip -\arraycolsep
  \let\@ifnextchar\new@ifnextchar
  \array{#1}}
\makeatother
\usepackage{blindtext}

\newcommand{\Real}{{\mathds R}} 

\newcommand{\Nat}{{\mathds N}}

\makeatletter
{}
{}
\newtheorem{proposition}{Proposition}{}
\newtheorem{problem}{Problem}{}
{}
{}
{}
{}
\begin{document}
\begin{frontmatter}

\title{Immersion and Invariance-based Coding for Privacy in Remote Anomaly Detection} 

\author[First]{Haleh Hayati} 
\author[First]{Nathan van de Wouw}
\author[First]{Carlos Murguia} 

\address[First]{Department of Mechanical Engineering, Dynamics and Control Group, Eindhoven University of Technology, The Netherlands, \\
Emails: h.hayati@tue.nl, n.v.d.wouw@tue.nl, c.g.murguia@tue.nl.}

\begin{abstract}                
We present a framework for the design of coding mechanisms that allow remotely operating anomaly detectors in a privacy-preserving manner. We consider the following problem setup. A remote station seeks to identify anomalies based on system input-output signals transmitted over communication networks. However, it is not desired to disclose true data of the system operation as it can be used to infer private information. To prevent adversaries from eavesdropping on the network or at the remote station itself to access private data, we propose a privacy-preserving coding scheme to distort signals before transmission. As a next step, we design a new anomaly detector that runs on distorted signals and produces distorted diagnostics signals, and a decoding scheme that allows extracting true diagnostics data from distorted signals without error. The proposed scheme is built on the synergy of matrix encryption and system Immersion and Invariance (I\&I) tools from control theory. The idea is to immerse the anomaly detector into a higher-dimensional system (the so-called target system). The dynamics of the target system is designed such that: the trajectories of the original anomaly detector are immersed/embedded in its trajectories, it works on randomly encoded input-output signals, and produces an encoded version of the original anomaly detector alarm signals, which are decoded to extract the original alarm at the user side.
We show that the proposed privacy-preserving scheme provides the same anomaly detection performance as standard Kalman filter-based chi-squared anomaly detectors while revealing no information about system data.
\end{abstract}
\end{frontmatter}
\section{Introduction}
Scientific and technological advancements in today's hyperconnected world have resulted in a massive amount of user data being collected and processed by hundreds of companies over public networks. Companies use this data to provide personalized services and targeted advertising. However, these technologies have come at the cost of a significant loss of privacy in society. Depending on their resources, adversaries can infer sensitive (private) information about users/systems from public data available on the internet and/or unsecured networks/servers. That is why researchers from different fields (e.g., information theory, computer science, and control engineering) have been drawn to the broad research subject of privacy and security of Cyber-Physical Systems (CPSs), see \cite{Jerome1}, \cite{Takashi_3}, \cite{Carlos_Iman1}, \cite{ferrari2021safety}.\\
In many engineering applications, information about the system operation is obtained through sensor measurements. This data is then sent to remote stations through communication networks for signal processing and decision-making purposes. One of these applications is remote anomaly detection, where a remote station seeks to identify anomalies based on system input-output signals transmitted over communication networks. If the communication network is public/unsecured and/or the remote station is not trustworthy, adversaries might access and infer system sensitive data, e.g., the system state/configuration, reference signals, product specifications, etc. Therefore, disclosing true data of the system operation is not desired as it can be used to infer private information.\\
In recent years, various privacy-preserving schemes have been implemented to address privacy leakage in dynamical systems applications. Most of them rely on two different strategies for providing privacy: perturbation-based schemes and cryptographic techniques, see \cite{farokhi2020privacy}. Perturbation-based schemes inject randomness to maintain privacy, and cryptographic algorithms limit access to data through encryption.\\
Perturbation-based techniques, usually rely on information-theoretic measures, see \cite{Farokhi1}, \cite{hayati2021finite}, or differential privacy, see \cite{Jerome1}, \cite{Dwork}, \cite{ferrari2021differentially}. In these techniques, random noise from a known distribution is injected into sensitive data before sending it to the remote station. However, in these methods, the induced distortion is never removed, which leads to performance degradation.\\ 
In most cryptography methods, data is encrypted before transmitting to the remote station and then decrypted at the remote station before processing, see \cite{wan2018physical}. This technique is suitable for making eavesdropping attacks difficult over communication networks. However, it is still vulnerable to insider attacks. Using encryption methods like homomorphic encryption (HE), which do not require data to be decrypted before processing, can reduce the occurrence of insider attacks. HE allows computations over plain data by performing appropriate computations on the encrypted data, see \cite{Gatsis_homo}. However, standard HE methods (e.g., Paillier encryption, see \cite{paillier1999public}) work over rings of integers which makes reformulating algorithms to work for dynamical systems over infinite time horizons difficult, see \cite{murguia2020secure}. Even when this reformulation is possible, mapping algorithms working on the reals to operate on finite rings leads to performance degradation and large computational overhead, see \cite{kim2022comparison}.\\ 
To summarize, although current solutions improve privacy of dynamical systems, they often do it at the expense of data utility and system efficiency. Balancing these tradeoffs is a key challenge when implementing privacy solutions for dynamical systems. Novel privacy-preserving schemes must be designed to provide strict privacy guarantees with a fair computational cost and without compromising application performance.\\
The aim of this work is to devise synthesis tools that allow designing coding mechanisms that protect private information and allow the implementation of remote anomaly detection procedures -- here, we consider standard Kalman filter-based chi-squared change detection procedures, see \cite{murguia2019model}. 
We propose a privacy-preserving anomaly detection scheme built on the synergy of random coding and system Immersion and Invariance (I\&I) tools, see \cite{astolfi2003immersion}, from control theory. The main idea is to treat the standard anomaly detection algorithm as a dynamical system that we seek to immerse into a higher-dimensional system (the so-called target system). The dynamics of the target system must be designed such that: 1) trajectories of the standard anomaly detector are immersed/embedded in its trajectories; and 2) it operates on encoded data. We use random coding, which is formulated at the user side as a random change of coordinates that maps original private data to a higher-dimensional space. Such coding enforces that the target system produces an encoded version of the standard anomaly detector alarm signals. The encoded anomaly detector alarms are decoded at the user side using the inverse of the encoding map (so this is basically a homomorphic encryption scheme that operates over the reals). The proposed framework provides the same performance as the standard anomaly detector (i.e., when no coding is employed to protect against data inference), reveals no information about private data, and is computationally efficient. To the best of our knowledge, this is the first piece of work that provides a high level of privacy for remote anomaly detectors without affecting performance.\\
Summarizing, the main contribution of the paper is the development of a synthesis framework that allows to jointly design of coding schemes to randomize disclosed data and new anomaly detectors that work on encrypted data, does not lead to performance degradation, and provides unconditionally secure privacy guarantees (\cite{wang2008book}). We have explored the use of similar ideas for privacy in Federated Learning, see \cite{hayati2022privacy}.\\
To the best of the authors' knowledge, only \cite{ferrari2021differentially} addresses a similar problem setup (using very different techniques, though, and proving different guarantees). For remote anomaly detection in stochastic dynamical systems, the authors in \cite{ferrari2021differentially} consider differential privacy between driving inputs and distorted disclosed data as their notion of privacy and provide tools to select the variance of i.i.d. noise to be injected such that disclosed data guarantees differential privacy. Their results consider privacy at every time-step, i.e., they do not consider information disclosed over a history of observations. While this indeed gives privacy guarantees, it is well known that for dynamical systems, the more data is collected, the more accurate the input/state estimation can be. 
 Therefore, if the privacy scheme is to avoid input/state estimation, giving guarantees at every time-step is certainly not sufficient. 
\subsection{Notation}
The symbol $\Real$ stands for the real numbers, and $\Real^+$ denotes the set of positive real numbers. The symbol $\Nat$ stands for the set of natural numbers. The Euclidian norm in $\Real^n$ is denoted by $||X||$, $||X||^2=X^{\top}X$, where $^{\top}$ denotes transposition. The $n \times n$ identity matrix is denoted by $I_n$ or simply $I$ if $n$ is clear from the context. Similarly, $n \times m$ matrices composed of only ones and only zeros are denoted by $\mathbf{1}_{n \times m}$ and $\mathbf{0}_{n \times m}$, respectively, or simply $\mathbf{1}$ and $\mathbf{0}$ when their dimensions are clear. 
The notation $x\sim \mathcal{N}[\mu^x,\Sigma^x]$ means $x \in \Real^{n}$ is a normally distributed random vector with mean $E[x] = \mu^x \in \Real^{n}$ and covariance matrix $E[(x-\mu^x)(x-\mu^x)^T] = \Sigma^x \in \Real^{n \times n}$, where $E[a]$ denotes the expected value of the random vector $a$. 
 The left inverse of the matrix $X$ is denoted by $X^L$ where $X^L X = I$. 
 The operator $\operatorname{ker}[\cdot]$ denotes the kernel. 
\section{System Description and Problem Formulation}
\subsection{System Description}
We consider discrete-time stochastic systems of the form:
\begin{eqnarray}\label{systemdescryption}
\left\{ \begin{aligned}
x_{k+1} &= Ax_k + Bu_k + t_k + D \delta_k,\\
y_k &= Cx_k + w_k + F \delta_k,
\end{aligned} \right.
\end{eqnarray}
with time-index $k \in \Nat$, state $x_k \in {\mathbb{R}^{{n_x}}}$, measurable output $y_k \in {\mathbb{R}^{{n_y}}}$, and \emph{known} input $u_k \in {\mathbb{R}^{{n_u}}}$. Vector $\delta_k \in {\mathbb{R}^{{n_\delta}}}$ models changes in the system dynamics due to faults or anomalies (thus, in the absence of anomalies $\delta_k=\mathbf{0}$). Matrices $(A,B,C,D,F)$ are of appropriate dimensions, and the pair $(A,C)$ is detectable. The state disturbance $t_k$ and the output disturbance $w_k$ are multivariate i.i.d. Gaussian processes with zero mean and positive definite covariance matrices ${\Sigma ^t}$ and ${\Sigma ^w}$, respectively. The initial state $x_1$ is assumed to be a Gaussian random vector with $E[x_1]=\mu_1^x \in \mathbb{R}^{n_x}$ and covariance matrix $\Sigma^x_1 \in \mathbb{R}^{n_x \times n_x}$, $\Sigma^x_1 > 0$. Disturbances $t_k$ and $w_k$ and the initial condition $x_1$ are mutually independent. We assume that matrices (vectors) $(A,B,C,\Sigma^x_1,\mu^x_1,\Sigma^t,\Sigma^w)$ and input $u_k$ are known for all $k$. 
\subsection{Anomaly Detection}
The aim of the anomaly detection algorithm is to identify anomalies (i.e., when $\delta_k \neq \mathbf{0}$ in \eqref{systemdescryption}) using system input-output signals. Here, we consider standard Kalman filter-based chi-squared change detection procedures, see \cite{murguia2019model}. The main idea is to use an estimator to anticipate the system evolution in the absence of anomalies. This prediction is subsequently compared with actual measurements from sensors. If the difference between what is measured and the estimation (commonly referred to as residuals) is larger than expected, there might be an anomaly in the system. We use steady-state Kalman filters as the state estimator.
\subsubsection{Steady-state Kalman Filter (anomaly-free case):}
Consider the following one step-ahead Kalman filter, see \cite{Astrom}, for \eqref{systemdescryption}:
\begin{equation}\label{eq5b}
\hat{x}_{k+1} = A \hat{x}_k + Bu_k + L (y_k - C\hat{x}_k),
\end{equation}
with estimated state $\hat{x}_k \in {\mathbb{R}^{{n_x}}}$, $\hat{x}_1 = E[x_1]$, and output injection gain matrix $L \in {\mathbb{R}^{{n_x} \times {n_y}}}$. Define the estimation error $e_k := x_k - \hat{x}_k$. The optimal filter gain $L$ minimizing the trace of the asymptotic estimation error covariance matrix $P := \lim_{k \rightarrow \infty} E[e_k e_k^\mathrm{T}]$ in the absence of anomalies is given by
\begin{equation}
    L:=\left(A P C^\top\right)\left(\Sigma^w + C P C^\top\right)^{-1},
\end{equation}
where $P$ is the solution of the Riccati equation:
\begin{equation}
    A P A^\top-P+ \Sigma^t=A P C^\top\left(\Sigma^w +C P C^\top\right)^{-1} C P A^\top.\label{solveP}
\end{equation}
Equation \eqref{solveP} always has a unique solution $P$ under the assumption of detectability of the pair $(A,C)$, see \cite{Astrom}.
\subsubsection{Residuals and Hypothesis Testing:}
Consider the process dynamics \eqref{systemdescryption} and the steady-state Kalman filter \eqref{eq5b}, and define the residual sequence $r_k := y_k - C\hat{x}_k$, which can be shown to evolve according to the difference equation:
\begin{align}\label{eq7b}
\left\{
\begin{aligned}
    e_{k+1} &= (A-L C) e_k + t_k - L w_k + (D-LF) \delta_k, \\
    r_k &= C e_k + w_k + F \delta_k.
\end{aligned}
\right.
\end{align}
If there are no anomalies ($\delta_k=\mathbf{0}$), the mean and the covariance matrix of the residual are as follows
\begin{align}
\left\{
\begin{aligned}
    &E[r_{k}] = C E[e_k] + E[w_k]= \mathbf{0}_{n_y \times 1}, \\[2mm]
    &E[r_{k} r^{\top} _{k}]=CPC^\top +\Sigma^w =: \Sigma \in {\mathbb{R}^{{n_y} \times {n_y}}}.
\end{aligned}\label{eq9}
\right.
\end{align}
For this residual, we identify two hypotheses to be tested: $\mathcal{H}_0$ the normal mode (no anomalies) and $\mathcal{H}_1$ the faulty mode (with anomalies). Under the normal mode, $\mathcal{H}_0$, the statistics of the residual are given by \eqref{eq9}.
However, in the presence of anomalies ($\mathcal{H}_1$), we expect that the statistics of the residual are different from those in the normal mode, i.e.,
\begin{equation}\label{eq10}
\mathcal{H}_1:\left\{
\begin{aligned}
    &E[r_{k}] \ne  \mathbf{0}_{n_y \times 1}, \text{ and/or}\\[2mm]
    &E[r_{k} r^{\top} _{k}] \ne \Sigma.
\end{aligned}
\right.
\end{equation}

There exist many well-known hypothesis testing techniques that could be used to examine the residual and subsequently detect anomalies/faults, for instance, (CUSUM) (\cite{murguia2016characterization}), generalized likelihood ratio testing (\cite{basseville1988detecting}), 
Compound Scalar Testing (CST) (\cite{gertler1988survey}), etc. Each technique has its own advantages and disadvantages, depending on the scenario. In this manuscript, we consider a particular case of CST, namely the so-called chi-squared change detection procedure, see \cite{murguia2019model}.
\subsubsection{Distance Measure and Chi-Squared Procedure:}
The input to the chi-squared procedure is a distance measure $z_k \in {\mathbb{R}}$, i.e., a measure of how deviated the estimator is from the sensor measurements. Here, we use the quadratic (in the residual) distance measure:
\begin{equation} \label{distance}
    z_k = r^\top _k \Sigma^{-1} r_k,
\end{equation}
where $r_k$ and $\Sigma$ are the residual sequence and its covariance matrix introduced in \eqref{eq9}. If there are no anomalies ($\delta_k=\mathbf{0}$), $E[r_{k}] = \mathbf{0}$ and $E[r_{k} r^{\top} _{k}]= \Sigma$; it follows that
\begin{align}\label{eq11}
\left\{
\begin{array}{ll}
    E[z_{k}] &= \text{tr}[\Sigma ^{-1} \Sigma]+ E[r_k]^\top \Sigma ^{-1} E[r_k] = n_y,\\[2mm]
    \text{var}[z_{k}]&=  2\text{tr}[\Sigma ^{-1} \Sigma \Sigma ^{-1} \Sigma]+ 4E[r_k]^\top \Sigma ^{-1} \Sigma \Sigma ^{-1} E[r_k] \\[1mm]
    &= 2n_y,
\end{array}
\right.
\end{align}
see, e.g., \cite{ross2014introduction} for details. Because $r_k \sim  \mathcal{N}(\mathbf{0},\Sigma)$, $z_k$ follows a chi-squared distribution with $n_y$ degrees of freedom. The chi-squared procedure is characterized by $z_k$ and its cumulative distribution as follows:
\\[1mm]
\noindent\makebox[\linewidth]{\rule{\linewidth}{0.8pt}}
\textbf{Chi-Squared procedure:}\\
\begin{equation}
a_k=m\left(z_{k}\right):= \begin{cases}0 &\text{If}\,\,\, z_{k} \le \alpha, \\
1 &\text{If}\,\,\, z_{k}>\alpha. \end{cases}
\end{equation}
\textbf{Design parameter:} threshold $\alpha \in \mathbb{R}^{+}$. \\
\textbf{Output:} alarm signal $a_k$.\\
\noindent\makebox[\linewidth]{\rule{\linewidth}{0.8pt}}

The idea is that if $z_k$ exceeds the threshold $\alpha$, alarms are triggered, and the time instant $k$ is identified as an alarm time. The parameter $\alpha$ is selected to satisfy a desired false alarm rate $\mathcal{A}^*$. Assume that there are no anomalies/faults and consider the chi-squared procedure with threshold $\alpha \in \mathbb{R^+}$, and $r_k \sim  \mathcal{N}(
\mathbf{0},\Sigma)$. Let $\alpha=\alpha^{*}:=2 \mathcal{P}^{-1}\left(\frac{n_y}{2}, 1-\mathcal{A}^{*}\right)$, where $\mathcal{P}^{-1}$ denotes the inverse regularised lower incomplete gamma function, then the false alarm rate $\mathcal{A}$, induced by the threshold $\alpha^*$, satisfies $\mathcal{A} = \mathcal{A}^*$ (see \cite{murguia2019model} for more details).

Summarizing, the complete anomaly detector is given as follows:
\begin{equation}\label{standarddetector}
\left\{
\begin{aligned}
\hat{x}_{k+1}
&=A \hat{x}_{k}+B u_{k}+L\left(y_{k}-C \hat{x}_{k}\right), \\
r_{k}&=y_{k}-C \hat{x}_{k},\\
z_{k}&=r_{k}^{\top} \Sigma r_{k}, \\
a_k&= \begin{cases}0 &\text{If}\,\,\, z_{k} \le \alpha, \\
1 &\text{If}\,\,\, z_{k}>\alpha. \end{cases}
\end{aligned}\right.
\end{equation}
Therefore, the anomaly detector at the remote station uses the system input $u_k$ and measurement $y_k$, which are transmitted over communication networks, to create the alarm signal $a_k$. Signal $a_k$ is sent back to the user through the networks. Note that information  about the system dynamics (state, model, references, etc.) can be inferred from the transmitted data. If the communication network is public/unsecured and/or the remote station is not a trusted party, information about the system dynamics needs to be fully disclosed to run the anomaly detector in \eqref{standarddetector}. To avoid this, in this work, transmitted data (input, measurement, and alarm signals) is considered private data that we aim to hide from adversaries by coding.
\subsection{Immersion Map and Target System}
The goal of the proposed privacy-preserving anomaly detection scheme is to make the inference of private data from the encoded disclosed data as hard as possible without degrading the performance of the anomaly detector. We seek to design a coding mechanism using system immersion and invariance tools from control theory.\\
System immersion refers to embedding the trajectories of a dynamical system into the trajectories of a different higher-dimensional system (the so-called target system), see \cite{astolfi2003immersion}. That is, there is a bijection between the trajectories of both systems (here referred to as the immersion map), and thus having a trajectory of the target system uniquely determines a trajectory of the original system via the immersion map.\\
In our setting, the idea is to immerse the dynamics of the standard anomaly detector \eqref{standarddetector} into a target dynamical system -- referred hereafter as the target anomaly detector. The dynamics of the target anomaly detector must be designed such that: 1) trajectories of the standard anomaly detector \eqref{standarddetector} are immersed in its trajectories via a known immersion map; 2) the target anomaly detector works on encoded input-measurement signals (here, we use random coding); and 3) the target anomaly detector produces an encoded version of the standard anomaly detector alarms that we can later decode at the user side. Once we have designed the target system and the immersion map, we can use them in real-time to operate on encoded input-measurement signals and to encode anomaly detector alarms. The user can decode the encoded alarm signal using the left-inverse of the encoding map.\\
Denote the state generated by the target Kalman filter as $\hat{x}'_k \in \mathbb{R}^{\tilde{n}_x}$ with ${\tilde{n}_x}>{n_x}$. 
Consider the following target anomaly detector:
\begin{equation}\label{targetdetector}
\text{Target anomaly detector}:
\left\{
\begin{aligned}
\hat{x}'_{k+1}&=\tilde{g}\left(\hat{x}_{k}^{\prime}, \tilde{u}_{k}, \tilde{y}_{k}\right), \\
\tilde{r}_{k}&=\tilde{h}\left(\tilde{y}_{k}, \hat{x}'_{k}\right),\\
\tilde{z}_{k}&=\tilde{f}\left(\tilde{r}_{k}, \tilde{y}_{k}\right),\\
\tilde{a}_k&=\tilde{m}\left(\tilde{z}_{k}, \tilde{y}_{k}\right),
\end{aligned}\right.
\end{equation}
with functions $\tilde{g}(\cdot)$, $\tilde{h}(\cdot)$, $\tilde{f}(\cdot)$, and $\tilde{m}(\cdot)$
to be designed, initial filter state $\hat{x}'_0$, and encoded input $\tilde{u}_{k}$, measurement $\tilde{y}_{k}$, residual $\tilde{r}_{k}$, distance measure $\tilde{z}_{k}$, and alarm signal $\tilde{a}_k$. We define encoding maps of input-measurement signals as $\tilde{y}_k=\pi_1(y_k)$ and $\tilde{u}_k=\pi_2(u_k)$ to be designed.\\
The standard anomaly detector in \eqref{standarddetector} is immersed in the target anomaly detector \eqref{targetdetector} through a function $\pi_3:\mathbb{R}^{n_x} \to \mathbb{R}^{\tilde{n}_x}$ to be designed satisfying:
\begin{equation}
\hat{x}_{k}^{\prime}=\pi_3\left(\hat{x}_{k}\right), \label{immersionmapping}
\end{equation}
for all $k$. We refer to this function $\pi_3(\cdot)$ as the \emph{immersion map}. To have \eqref{immersionmapping} for all $k$, the target system \eqref{targetdetector}, the function $\pi_3(\cdot)$, and the initial filter state $\hat{x}_{0}^{\prime}$ must be designed such that starting the target filter state from the manifold \eqref{immersionmapping}, i.e., $\hat{x}_{0}^{\prime}=\pi_3\left(\hat{x}_{0}\right)$, it always remain on this manifold. This leads to forward invariance condition under the dynamics of the standard Kalman filter in \eqref{standarddetector} and the target filter $\tilde{g}(\cdot)$. 
Defining an off-the-manifold error as $\epsilon_k=\hat{x}_{k}^{\prime}-\pi_3\left(\hat{x}_{k}\right)$, the manifold is forward invariant if the origin of its dynamics:
\begin{equation}\label{errordynamic}
\begin{aligned}
\epsilon_{k+1}&= \hat{x}'_{k+1}-\pi_3\left(\hat{x}_{k+1}\right)\\
&=\tilde{g}\left((\pi_3\left(\hat{x}_{k}\right) +\epsilon_k), \tilde{u}_k,\tilde{y}_k\right)-\pi_3\left(\hat{x}_{k+1}\right),
\end{aligned}
\end{equation}
is a fixed point, i.e., $\epsilon_k=\mathbf{0}$ implies $\epsilon_{k+1}=\mathbf{0}$. Substituting $\epsilon_k=\mathbf{0}$ and $\epsilon_{k+1}=\mathbf{0}$ into \eqref{errordynamic} we have:
\begin{equation}\label{errordynamic2}
\begin{aligned}
\tilde{g}\left(\pi_3\left(\hat{x}_{k}\right), \tilde{u}_k,\tilde{y}_k\right)=\pi_3\left(\hat{x}_{k+1}\right). 
\end{aligned}
\end{equation}
Hence, we need to enforce (by designing $\pi_3(\cdot)$, $\tilde{g}(\cdot)$, and $\hat{x}'_0$) that: \textbf{(a)} the initial condition of \eqref{targetdetector}, $\hat{x}'_0$, satisfies $\hat{x}'_0 = \pi_3(\hat{x}_0)$, which leads to $\epsilon_0=\mathbf{0}$ (start on the manifold); and \textbf{(b)} the dynamics of both algorithms match under the immersion map, i.e., \eqref{errordynamic2} is satisfied (invariance condition on the manifold). Therefore, using the expressions for $\hat{x}_{k+1}$ in \eqref{standarddetector}, the immersion map \eqref{immersionmapping}, and encoding maps, the invariance condition \eqref{errordynamic2} can be written as follows:
\begin{equation}\label{immersioncondition}
\begin{aligned}
&\tilde{g}\left(\pi_3\left(\hat{x}_{k}\right), \pi_2\left({u}_{k}\right), \pi_1\left({y}_{k}\right)\right)\\ &\hspace{20mm}= \pi_3\left(A \hat{x}_{k}+B u_{k}+L\left(y_{k}-C \hat{x}_{k}\right)\right), 
\end{aligned}
\end{equation}
for all $\hat{x}_k \in \mathbb{R}^{n_x}$. We refer to this equation as the \emph{immersion condition}.\\
In the target anomaly detector \eqref{targetdetector}, a distorted filter is used to compute encoded residuals, distance measures, and alarms, and then, the encoded alarms are sent back to the user for decoding. Therefore, the alarm signal must satisfy $\tilde{a}_k = \pi_4({a}_k)$ for some left-invertible mapping $\pi_4(\cdot)$ to be designed. The latter imposes an extra condition on encoding maps since to retrieve $a_k$ from $\tilde{a}_k$: \textbf{(c)} there must exist a function $\pi_4^L:\mathbb{R}^{\tilde{n}_a} \to \mathbb{R}$ satisfying the following left-invertibility condition:
\begin{equation} \label{left_inverse}
\pi_4^L \left( \tilde{a}_k \right) = a_k.
\end{equation}
If such $\pi_4^L(\cdot)$ and $\pi_4(\cdot)$ exist, the user can retrieve the original alarm signal by passing the encoded one through function $\pi_4^L(\cdot)$. 
We now have all the machinery required to state the problem we seek to solve.\\[2mm]
\begin{problem}\emph{\textbf{(Privacy-Preserving Anomaly Detector)}} Consider the standard anomaly detector \eqref{standarddetector} and the target anomaly detector \eqref{targetdetector}. Design encoding maps, $\pi_1(\cdot)$, $\pi_2(\cdot)$, and $\pi_4(\cdot)$, the immersion map $\pi_3(\cdot)$, and functions $\tilde{g}(\cdot)$, $\tilde{h}(\cdot)$, $\tilde{f}(\cdot)$, and $\tilde{m}(\cdot)$ in \eqref{targetdetector} such that: \textbf{(a)} the initial condition of estimated state satisfies $\hat{x}'_0 = \pi_3(\hat{x}_0)$ (start on the manifold); \textbf{(b)} the dynamics of both algorithms match under the immersion map, i.e., the immersion condition \eqref{immersioncondition} is satisfied (invariance condition on the manifold); and \textbf{(c)} there exists a function $\pi_4^L(\cdot)$ satisfying \eqref{left_inverse} (left invertability condition).
\end{problem}
\section{Solution to Problem 1}
In this section, we construct a class of the proposed target anomaly detection algorithms by deriving particular expressions for all functions in Problem $1$. We start with function $\tilde{g}(\cdot)$ and the immersion condition \eqref{immersioncondition}. Let the coding maps $\pi_1(\cdot)$ and $\pi_2(\cdot)$, and the immersion map $\pi_3(\cdot)$ be affine functions as follows:
\begin{equation}\label{privacymechanism2}
    \left\{\begin{array}{l}
    \tilde{y}_k=\pi_1(y_k):=\Pi_1 y_k+b^1_k, \\
    \tilde{u}_k=\pi_2(u_k):=\Pi_2 u_k+b^2_k,\\
    \hat{x}'_k=\pi_3(\hat{x}_k):=\Pi_3 \hat{x}_k,
    \end{array}\right.
\end{equation}
for some $\Pi_1 \in \mathbb{R}^{\tilde{n}_y \times n_y}$, $\Pi_2 \in \mathbb{R}^{\tilde{n}_u \times n_u}$, $\Pi_3 \in \mathbb{R}^{\tilde{n}_x \times n_x}$, with $\tilde{n}_y> n_y$, $\tilde{n}_u> n_u$, $\tilde{n}_x> n_x$, and $b^1_k \in \mathbb{R}^{\tilde{n}_y}$, and $b^2_k \in \mathbb{R}^{\tilde{n}_u}$ -- with slight abuse of notation, we let $b^1_k$ and $b^2_k$ change with $k$ independently of $y_k$ and $u_k$. Then, the immersion condition \eqref{immersioncondition} reduces to
\begin{equation}\label{immersioncondition2}
    \begin{aligned}
    &\tilde{g}\left(\Pi_3 \hat{x}_k, \Pi_2 u_k+b^2_k,\Pi_1 y_k+b^1_k\right)\\&\hspace{20mm}=\Pi_{3} \left( A \hat{x}_{k}+B u_{k}+L\left(y_{k}-C \hat{x}_{k}\right)\right).
\end{aligned}
\end{equation}
Let the function $\tilde{g}(\cdot)$ be of the form 
\begin{equation}\label{eq:20}
    \begin{aligned}
    &\tilde{g}\left(\hat{x}_{k}^{\prime}, \tilde{u}_{k}, \tilde{y}_{k}\right)\\ &\hspace{10mm}:=\Pi_3\left( A M_3 \hat{x}_{k}^{\prime}+B M_2 \tilde{u}_{k}+L\left(M_1\tilde{y}_{k}-C M_3 \hat{x}_{k}^{\prime}\right)\right),
\end{aligned}
\end{equation}
for some $M_1 \in \mathbb{R}^{n_y \times \tilde{n}_y}$, $M_2 \in \mathbb{R}^{n_u \times \tilde{n}_u}$, and $M_3 \in \mathbb{R}^{n_x \times \tilde{n}_x}$ to be designed. Hence, the immersion condition \eqref{immersioncondition2} takes the form
\begin{equation}\label{eqc1}
\begin{aligned}
&\Pi_3 \left( A M_3 \hat{x}_{k}^{\prime}+B M_2 \tilde{u}_{k}+L\left(M_1\tilde{y}_{k}-C M_3 \hat{x}_{k}^{\prime}\right)\right)\\ &\hspace{30mm}=\Pi_{3} \left( A \hat{x}_{k}+B u_{k}+L\left(y_{k}-C \hat{x}_{k}\right)\right). 
\end{aligned}
\end{equation}
Note that the choice of $\tilde{g}$ in \eqref{eq:20} provides a prescriptive design in terms of the original estimator dynamics. That is, we exploit the knowledge of the original dynamics and build the target system on top of it in an algebraic manner. To satisfy \eqref{eqc1}, we must have $M_1\left(\Pi_1 y_k +b^1_k\right)=y_k$, $M_2\left(\Pi_2 u_k +b^2_k\right)=u_k$, and $M_3\Pi_3 \hat{x}_k=\hat{x}_k$, i.e., $M_i \Pi_i=I$, for $i \in \{1,2,3\}$, $b^1_k \in \text{ker}[M_1]$, and $b^2_k \in \text{ker}[M_2]$. It follows that: 1) $\Pi_i$, $i \in \{1,2,3\}$, \emph{must be of full column rank} (i.e., $\text{rank}[\Pi_1]=n_y$, $\text{rank}[\Pi_2]=n_u$, and $\text{rank}[\Pi_3]=n_x$); 2) $M_i$ is a left inverse of $\Pi_i$, i.e.  $M_i = \Pi^L_i$ (which always exists given the rank of $\Pi_i$); and 3) $b^1_k \in \text{ker}[\Pi^L_1]$, $b^2_k \in \text{ker}[\Pi^L_2]$, and these kernels are always nontrivial because $\Pi_i$ is full column rank by construction. So the final form for $\tilde{g}(\cdot)$ in \eqref{targetdetector} is given as
\begin{equation}\label{eqc2}
\begin{aligned}
\tilde{g}\left(\hat{x}_{k}^{\prime}, \tilde{u}_{k}, \tilde{y}_{k}\right)&=\Pi_{3} ( A \Pi_{3}^{L} \hat{x}_{k}^{\prime}+B \Pi_{2}^{L} \tilde{u}_{k}\\&+L\left(\Pi_{1}^{L} \tilde{y}_{k}-C \Pi_{3}^{L} \hat{x}_{k}^{\prime}\right)).
\end{aligned}
\end{equation}
At every $k$, vectors $b^1_k$ and $b^2_k$ are designed to satisfy $\Pi_{1}^{L} b^1_k=0$ and $\Pi_{2}^{L} b^2_k=0$ and used to construct the coding maps in \eqref{privacymechanism2}. To increase privacy, we use these $b^1_k$ and $b^2_k$ to add randomness to the mappings by exploiting the nontrivial kernels of $\Pi_{1}^{L}$ and $\Pi_{2}^{L}$. Without loss of generality, we let them be of the form $b^1_k=N_1 s^1_k$ and $b^2_k=N_2 s^2_k$,
for some matrices $N_1 \in \mathbb{R}^{\tilde{n}_y \times (\tilde{n}_y - n_y)}$
and $N_2 \in \mathbb{R}^{\tilde{n}_u \times (\tilde{n}_u - n_u)}$ expanding the kernels of $\Pi_{1}^{L}$ and $\Pi_{2}^{L}$, respectively (i.e., $\Pi_{1}^{L}N_1=0$, and $\Pi_{2}^{L}N_2=0$) and some random vectors $s^1_k \in \mathbb{R}^{(\tilde{n}_y - n_y)}$ and $s^2_k \in \mathbb{R}^{(\tilde{n}_u - n_u)}$. Hence, we have $\Pi_{1}^{L} b^1_k=\Pi_{2}^{L} b^2_k=\mathbf{0}$ for all $k$ and $b^1_k=N_1 s^1_k$ and $b^2_k=N_2 s^2_k$ change randomly with $k$.

So far, we have designed the function $\tilde{g}(\cdot)$, the immersion map $\pi_3(\cdot)$, and encoding maps $\pi_1(\cdot)$ and $\pi_2(\cdot)$ to satisfy the immersion condition \eqref{immersioncondition}. Next, we seek for functions $\tilde{h}(\cdot)$, $\tilde{f}(\cdot)$, and $\tilde{m}(\cdot)$ such that the target anomaly detector \eqref{targetdetector} encodes $\tilde{r}_k$, $\tilde{z}_k$, and $\tilde{a}_k$ through affine mappings, and there exists a decoding function that extracts the true alarm signal ${a}_k$ from the encoded $\tilde{a}_k$ (the left-invertibility condition \eqref{left_inverse}).
\begin{figure*}
  \centering
\includegraphics[width=0.58\textwidth]{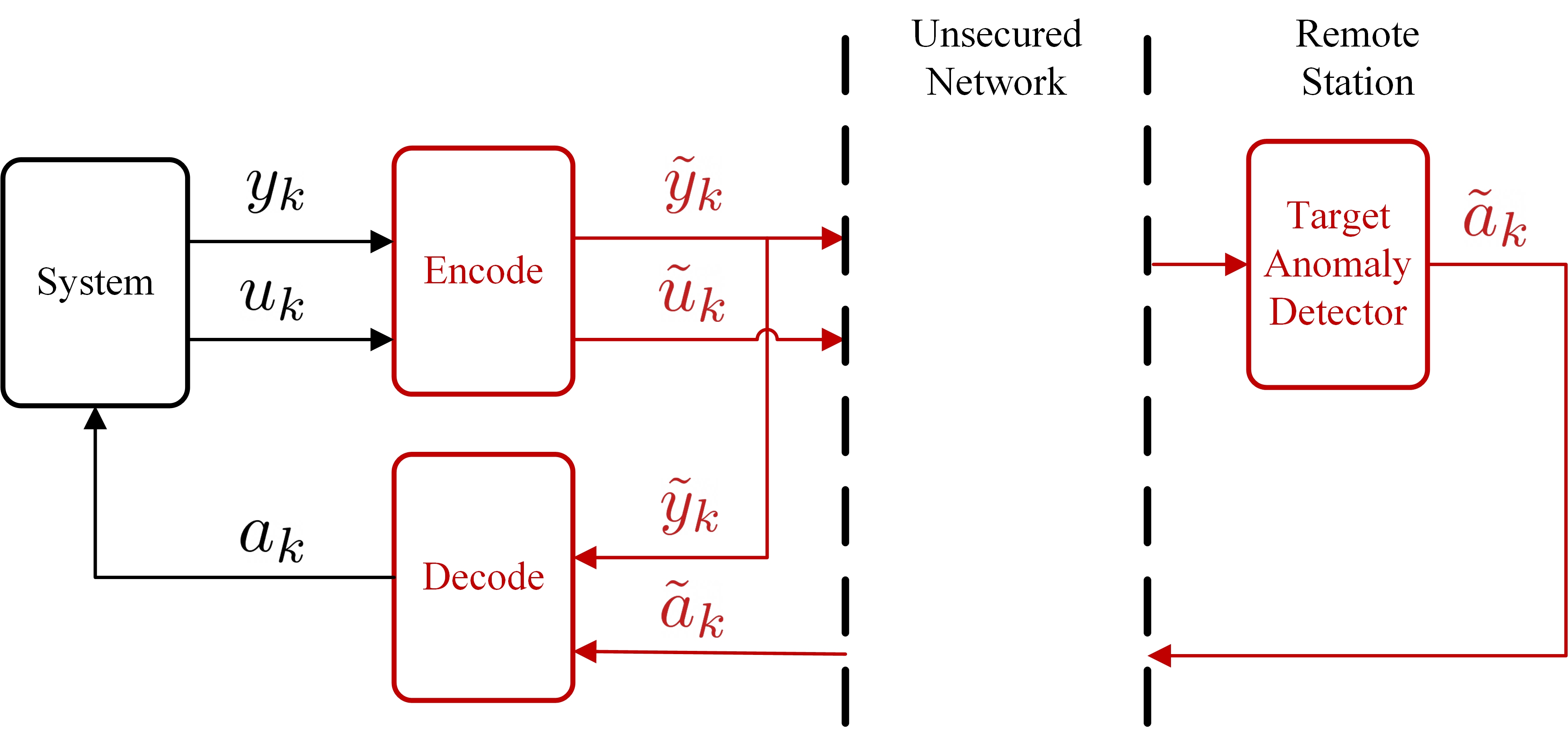}
    \caption{Encoding/decoding scheme.}
    \label{fig1}
\end{figure*}
Let $\tilde{h}(\cdot)$ be of the form $\tilde{h}\left(\tilde{y}_k, \hat{x}'_k\right)=\Pi_7 \tilde{y}_k - M_4 \hat{x}'_k$ for some $\Pi_7 \in \mathbb{R}^{\tilde{n}_r \times \tilde{n}_y}$ and $M_4 \in \mathbb{R}^{\tilde{n}_r \times \tilde{n}_x}$ to be designed. We aim to enforce that $\tilde{r}_k$ is an affine function of $r_k$, i.e.,  $\tilde{r}_k=\Pi_5 {r}_k + b^5_k$ for some matrix $\Pi_5$ and sequence $b_k^5$. Note that to satisfy the latter, we need to match this desired expression for $\tilde{r}_k$ with $\tilde{h}(\tilde{y}_k,\hat{x}'_k)$ above, given \eqref{privacymechanism2}. Then, we have
\begin{align}
    \Pi_7 \left( \Pi_1 y_k + b_k^1\right) - M_4 \left(\Pi_3 \hat{x}_k\right) = \Pi_5 (y_k - C\hat{x}_k) + b^5_k.\label{rkdesign}
\end{align}
To satisfy \eqref{rkdesign}, we require $\Pi_5 = \Pi_7 \Pi_1$, $M_4 = \Pi_7 \Pi_1 C \Pi_3^L$, and $b^5_k=\Pi_7 b^1_k$. Hence, $\tilde{h}(\cdot)$ is constructed as
\begin{equation}\label{rkf}
    \tilde{h}\left(\tilde{y}_k, \hat{x}'_k\right) := \Pi_7 (\tilde{y}_k - \Pi_1 C \Pi_3^L \hat{x}'_k), 
\end{equation}
and consequently, $\tilde{r}_k$ is an affine function of $r_k$ given by
\begin{equation}\label{rkdesign2}
    \tilde{r}_k=\Pi_7 \left( \Pi_1 r_k +  b_k^1 \right).
\end{equation}
Note that the only matrix to be designed with this choice for $\tilde{h}(\cdot)$ is matrix $\Pi_7$. We only require $\Pi_7$ to be of full column rank as we will use its left inverse in what follows. We follow the same reasoning for the design of $\tilde{f}(\cdot)$, i.e., we design it to make the encoded $\tilde{z}_k$ an affine function of $z_k$ given by $\tilde{z}_k=\Pi_6 {z}_k + b^6_k$. By matching this latter expression of $\tilde{z}_k$ with the affine maps in \eqref{privacymechanism2} and the above expression for $\tilde{r}_k$ in \eqref{rkdesign2} yields:
\begin{equation}\label{zkf}
    \tilde{f}\left(\tilde{r}_{k}, \tilde{y}_{k}\right) := \Pi_6 \left(\left(\Pi^L_1 \Pi^L_7 \tilde{r}_k\right)^{\top} \Sigma \left(\Pi^L_1 \Pi^L_7 \tilde{r}_k\right) \right)+ \Pi_8 \tilde{y}_{k}. 
\end{equation}
Hence, $b_k^6 := \Pi_8\tilde{y}_k$. The encoded $\tilde{y}_{k}$ is used in $\tilde{f}\left(\tilde{r}_{k}, \tilde{y}_{k}\right)$ to inject randomness into $\tilde{z}_k$ to increase privacy. Therefore, $\tilde{z}_k$ is an affine function of $z_k$ given by
\begin{equation}\label{zkdesign2}
    \tilde{z}_k=\Pi_6 z_k + \Pi_8 \tilde{y}_{k},
\end{equation}
for some \emph{full rank} $\Pi_6 \in \mathbb{R}^{\tilde{n}_z \times 1}$ and $\Pi_8 \in \mathbb{R}^{\tilde{n}_z \times \tilde{n}_y}$ to be designed. Using the same technique for the design of $\tilde{m}(\cdot)$, we have
\begin{equation}\label{distortedalarm}
\begin{aligned}
    \tilde{a}_k = \tilde{m}\left(\tilde{z}_k,\tilde{y}_k\right)&:=\Pi_4 m\left( \Pi^L_6 \left( \tilde{z}_k - \Pi_8 \tilde{y}_k\right) \right) + \Pi_9 \tilde{y}_k \\&= \Pi_4 a_k + \Pi_9 \tilde{y}_k,
    \end{aligned}
\end{equation}
for some \emph{full rank} $\Pi_4 \in \mathbb{R}^{\tilde{n}_a \times 1}$ and $\Pi_9 \in \mathbb{R}^{\tilde{n}_a \times \tilde{n}_y}$ to be designed and $b^4_k:=\Pi_9 \tilde{y}_k$. Matrix $\Pi^L_6$ always exists because $\Pi_6$ is full column rank.

Finally, we seek a function $\pi_4^L(\cdot)$ satisfying \eqref{left_inverse} (condition \textbf{(c)} in Problem $1$). Given \eqref{distortedalarm}, condition \eqref{left_inverse} is written as
\begin{align} \label{left_inverse2}
\pi_4^L \left( \Pi_4 a_k + \Pi_9  \tilde{y}_k\right) =  a_k,
\end{align}
which trivially leads to
\begin{equation}
    \pi_4^L(\tilde{a}_k,\tilde{y}_k) := \Pi_4^L \left( \tilde{a}_k-\Pi_9 \tilde{y}_k\right). \label{inversemap}
\end{equation}
Matrix $\Pi_4^L$ always exists due to the rank of $\Pi_4$.
\begin{proposition}\emph{\textbf{(Solution to Problem $1$)}} The encoding maps:
\begin{equation}\label{privacymechanisms}
    \left\{\begin{array}{l}
\tilde{y}_k=\Pi_1 y_k + N_1 s_k^1,\\
\tilde{u}_k=\Pi_2 u_k + N_2 s_k^2,\\
\tilde{a}_k = \Pi_4 a_k + \Pi_9 \tilde{y}_k,
    \end{array}\right.
\end{equation}
the target anomaly detector:
\begin{equation}\label{immersionsolution}
\left\{\begin{aligned}
\hat{x}'_{k+1}&=\tilde{g}\left(\hat{x}_{k}^{\prime}, \tilde{u}_{k}, \tilde{y}_{k}\right)\\&:=\Pi_{3} \left(A \Pi_3^L \hat{x}_k^{\prime} + B \Pi_2^L \tilde{u}_k+ L_k (\Pi_1^L \tilde{y}_k - C\Pi_3^L\hat{x}^{\prime}_k) \right), \\
\tilde{r}_{k}&=\tilde{h}\left(\tilde{y}_{k}, \hat{x}'_{k}\right):= \Pi_7 (\tilde{y}_k - \Pi_1 C \Pi_3^L \hat{x}'_k),\\
\tilde{z}_{k}&=\tilde{f}\left(\tilde{r}_{k}, \tilde{y}_{k}\right)\\&:=\Pi_6 \left( \Pi^L_1 \Pi^L_7 \tilde{r}_k \right)^T\Sigma^{-1}\left( \Pi^L_1 \Pi^L_7 \tilde{r}_k \right)+ \Pi_8 \tilde{y}_{k},\\
\tilde{a}_k&=\tilde{m}\left(\tilde{z}_{k}, \tilde{y}_{k}\right)\\&:= \begin{cases}\Pi_9 \tilde{y}_k &\text{If}\,\,\,\,\, \left( \Pi^L_6 \left( \tilde{z}_k - \Pi_8 \tilde{y}_k\right) \right) \le \alpha, \\
\Pi_4 + \Pi_9 \tilde{y}_k &\text{If}\,\,\,\,\, \left( \Pi^L_6 \left( \tilde{z}_k - \Pi_8 \tilde{y}_k\right) \right)>\alpha, \end{cases}
\end{aligned}\right.
\end{equation}

and inverse function:
\begin{equation}\label{inversesolution}
a_k=\pi_4^L(\tilde{a}_k,\tilde{y}_k) := \Pi_4^L \left( \tilde{a}_k-\Pi_9 \tilde{y}_k\right),
\end{equation}
with \emph{full column rank matrices} $\Pi_1 \in \mathbb{R}^{\tilde{n}_y \times n_y}$, $\Pi_2 \in \mathbb{R}^{\tilde{n}_u \times n_u}$, $\Pi_3 \in \mathbb{R}^{\tilde{n}_x \times n_x}$, $\Pi_4 \in \mathbb{R}^{\tilde{n}_a \times 1}$, $\Pi_6 \in \mathbb{R}^{\tilde{n}_z \times 1}$, and $\Pi_7 \in \mathbb{R}^{\tilde{n}_r \times \tilde{n}_y}$, and full rank matrices $\Pi_8 \in \mathbb{R}^{\tilde{n}_z \times \tilde{n}_y}$ and $\Pi_9 \in \mathbb{R}^{\tilde{n}_a \times \tilde{n}_y}$ with $\tilde{n}_a,\tilde{n}_z>1$, provide a solution to Problem $1$.
\end{proposition}
\emph{\textbf{Proof}}: The proof follows from the analysis provided in the solution section above, Section $3$.
\hfill $\blacksquare$\\[2mm]
Based on Problem $1$ and the solution in Proposition $1$, there are no conditions on coding matrices $\Pi_i$, $i \in \{1,2,\dots,9\}$. Moreover, the random processes $s_k^1$ and $s^2_k$ are arbitrary and potentially unbounded. Therefore, they can be chosen to maximize privacy. Since this algorithm eliminates the encoding-induced distortion in the decoding step, we maximize privacy without sacrificing the anomaly detection performance. We maximize privacy by properly selecting the encoding matrices and random processes.\\
Intuitively, from \eqref{privacymechanisms} and \eqref{immersionsolution}, it is apparent that by making $\Pi_1$, $\Pi_2$, $\Pi_3$, $\Pi_4$, $\Pi_6$ and $\Pi_7$ small, and $b_k^1$, $b_k^2$, $\Pi_8$, and $\Pi_9$ large, the encoded data $\tilde{y}_k$, $\tilde{u}_k$, $\tilde{r}_k$, $\tilde{z}_k$, and $\tilde{a}_k$ `converge' to their random terms $b^1_k$, $b^2_k$, $\Pi_7 b_k^1$, $\Pi_8 b^1_k$, and $\Pi_9 b^1_k$, and diverge from the actual signals ${y}_k$, ${u}_k$, $r_k$, $z_k$, and $a_k$, respectively. Subsequently, inferring information about private data from disclosing data $\tilde{y}_k$, $\tilde{u}_k$, and $\tilde{a}_k$ would be more difficult for adversaries.\\
The flowchart of the proposed scheme is shown in Figure \ref{fig1}. As can be seen in this figure, all the shared information through the communication network ($\tilde{y}_k$, $\tilde{u}_k$, and $\tilde{a}_k$) between the user and the remote anomaly detector is encoded.
\section{General Guidelines for Implementation}
The summary of the implementation procedure of the proposed algorithm is as follows:

\noindent\makebox[\linewidth]{\rule{\linewidth}{0.8pt}}
\textbf{Synthesis and Operation Procedure:}\\
\begin{itemize}
    \item Randomly select `small' full rank matrices $\Pi_1$-$\Pi_4$, and $\Pi_6$-$\Pi_7$ of appropriate dimensions (see Proposition $1$).\vspace{2mm}
    \item Randomly select `large' full rank matrices $\Pi_8$-$\Pi_9$ of appropriate dimensions (see Proposition $1$).\vspace{2mm}
    \item Compute bases $N_1$ and $N_2$ of the kernels of $\Pi_{1}^{L}$ and $\Pi_{2}^{L}$, respectively.\vspace{2mm}
    \item Fix the distributions of the random processes $s_k^1$ and $s_k^2$ in \eqref{privacymechanisms}.\vspace{2mm}
    \item At every time step $k \in \mathbb{N}$, encode $y_k$ and $u_k$ according to \eqref{privacymechanisms} at the user side and send $\tilde{y}_k$ and $\tilde{u}_k$ to the remote station.\vspace{2mm}
    \item Have the remote station run the target anomaly detector in \eqref{immersionsolution} using the received encoded $\tilde{y}_k$ and $\tilde{u}_k$.\vspace{2mm}
    \item The remote station sends $\tilde{a}_k$ back to the user.\vspace{2mm}
    \item The user decodes the encoded alarm signal using the inverse function in \eqref{inversesolution}.
\end{itemize}
\noindent\makebox[\linewidth]{\rule{\linewidth}{0.8pt}}\\
\section{Security Analysis}
In the proposed scheme, because private data (input and measurement signals and anomaly detection alarms) is encoded in all iterations, adversaries can only obtain distorted data. 
Therefore, adversaries need to break the coding system to infer information about private data. Since the coding keys are random and changed at each time step (random additive terms), even if they manage to break some coding matrices at some time steps, they cannot infer the actual data due to the inclusion of randomness.\\
An unconditionally secure cryptosystem is a system that is secure against adversaries with unlimited computing resources and time, see \cite{diffie2019new}. In \cite{shannon1949communication}, Shannon proves that a necessary condition for an encryption method to be unconditionally secure is that the uncertainty of the secret key is larger than or equal to the uncertainty of the plaintext, see \cite{wang2008book}. He proposes a one-time pad encryption scheme in which the key is randomly selected and never used again. The one-time pad gives unbounded entropy of the key space, i.e., infinite key space, which provides unconditional security. Unconditional secrecy is lost when the key is not random or if it is reused. The unconditional security idea can be extended to the proposed coding mechanism. Since in the proposed I\&I-based privacy-preserving anomaly detection, the encoding keys are random and only used once, it provides infinite key space, and thus, it can be considered unconditionally secure.
\section{Illustrative Case Study}
The authors of \cite{murguia2019model} studied the anomaly detection problem for a well-stirred chemical reactor with a heat exchanger. We use this case study to demonstrate our results. The state, input, and output vectors of the considered reactor are:%
\begin{align*}
\left\{
\begin{array}{ll}
x_k =\begin{pmatrix} C_0\\T_0\\T_w\\T_m\end{pmatrix},
u_k =\begin{pmatrix} C_u\\T_u\\T_{w,u}\end{pmatrix},
y_k =\begin{pmatrix} C_0\\T_0\\T_w\end{pmatrix},
\end{array}
\right.
\end{align*}
where
\begingroup\makeatletter\def\f@size{9.0}\check@mathfonts
\def\maketag@@@#1{\hbox{\m@th\normalsize\normalfont#1}}%
\begin{align*}
\left\{
\begin{array}{ll}
C_0&: \text{Concentration of the chemical product},\\
T_0&: \text{Temperature of the product},\\
T_w&: \text{Temperature of the jacket water of the heat exchanger},\\
T_m&: \text{Coolant temperature},\\
C_u&: \text{Inlet concentration of the reactant}.\\
T_u&: \text{Inlet temperature},\\
T_{w,u}&: \text{Coolant water inlet temperature}.
\end{array}
\right.
\end{align*}
\begin{table*}
\noindent\rule{\hsize}{1pt}
\begin{equation} \label{eq:experiment}
\left\{\begin{array}{l}
A=\left(\begin{array}{cccc}
0.8353 & 0 & 0 & 0 \\
0 & 0.8324 & 0 & 0.0031 \\
0 & 0.0001 & 0.1633 & 0 \\
0 & 0.0280 & 0.0172 & 0.9320
\end{array}\right), \quad B=\left(\begin{array}{ccc}
0.0458 & 0 & 0 \\
0 & 0.0457 & 0 \\
0 & 0 & 0.0231 \\
0 & 0.0007 & 0.0006
\end{array}\right), \quad C=\left(\begin{array}{cccc}
1 & 0 & 0 & 0 \\
0 & 1 & 0 & 0 \\
0 & 0 & 1 & 0
\end{array}\right), \quad \Sigma^w = 0.001I_{n_y},\\
\Sigma^t=0.001I_{n_x},\quad L =\left(\begin{array}{ccc}
0.8271 & 0 & 0 \\
0 & 0.8243 & 0.0002 \\
0 & 0.0002 & 0.1619 \\
0 & 0.0481 & 0.0543
\end{array}\right),  \quad \Sigma=\left(\begin{array}{ccc}
1.0169 & 0 & 0 \\
0 & 1.0169 & 0.0001 \\
0 & 0.0001 & 1.0105
\end{array}\right),\quad D=\left(\begin{array}{cccc}
1\\2\\3\\4
\end{array}\right),\quad F=\left(\begin{array}{cccc}
1\\2\\3
\end{array}\right).
\end{array}\right.
\end{equation}
\noindent\rule{\hsize}{1pt}
\end{table*}
We use the discrete-time dynamics of the reactor introduced in \cite{murguia2021privacy} for our simulation-based case study with matrices as given in \eqref{eq:experiment}, normally distributed initial condition $x_1 \sim \mathcal{N}[(6.94;13.76;1;1)^\top,0.001 I]$, and reference signal $u_k = 50\cos[0.5k]^2$. For anomaly detection variables, we design the desired false alarm rate $\mathcal{A}^*=0.1$, and the induced threshold $\alpha^*=6.2514$ satisfying $\mathcal{A} = \mathcal{A}^*$. For the dimensions of the target anomaly detector algorithm matrices, we consider $\tilde{n}_x=8$, $\tilde{n}_y=4$, $\tilde{n}_u=4$, and $\tilde{n}_a=2$. Also, for designing target anomaly detector algorithm matrices, we randomly select small full rank matrices $\Pi_1$-$\Pi_4$ and $\Pi_6$-$\Pi_7$, and large full rank matrices $\Pi_8$-$\Pi_9$ based on the designed dimensions. Then, we compute bases $N_1$ and $N_2$ of the kernels of $\Pi_{1}^{L}$ and $\Pi_{2}^{L}$, respectively. The random processes $s_k^1$ and $s_k^2$ are defined as multivariate Gaussian variables with large mean and covariances.\\
First, in Figure \ref{yyp}, we show the effect of the proposed coding mechanisms, where we contrast actual and encoded measurement data. As can be seen in this figure, since the encoded measurement $\tilde{y}_k$ is based on random term $b_k^1$ for very small $\Pi_1$ and large $s^1_k$ in \eqref{privacymechanisms}, it is completely different from actual measurement $y_k$. Besides, the dimension of the measurement vector $y_k$ is increased from three to four in the encoded measurement $\tilde{y}_k$. Therefore, adversaries can not even access the actual dimension of the measurement vector. The same result is achieved for the comparison between actual and encoded input $u_k$ and $\tilde{u}_k$.\\ 
\begin{figure*}[t]
  \centering
  \subcaptionbox{}[.48\linewidth][c]{%
    \includegraphics[width=1\linewidth]{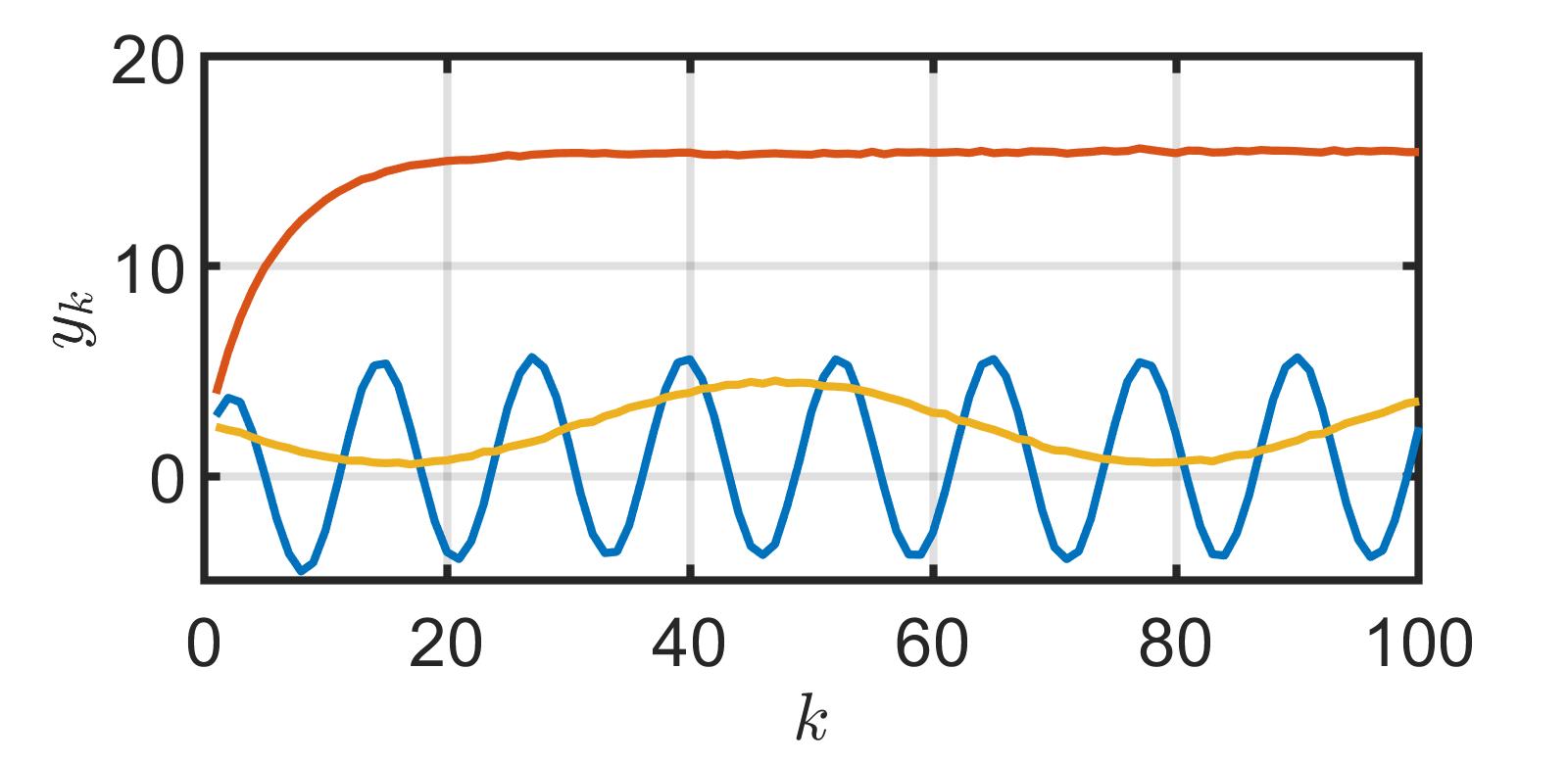}}\quad
  \subcaptionbox{}[.48\linewidth][c]{%
    \includegraphics[width=1\linewidth]{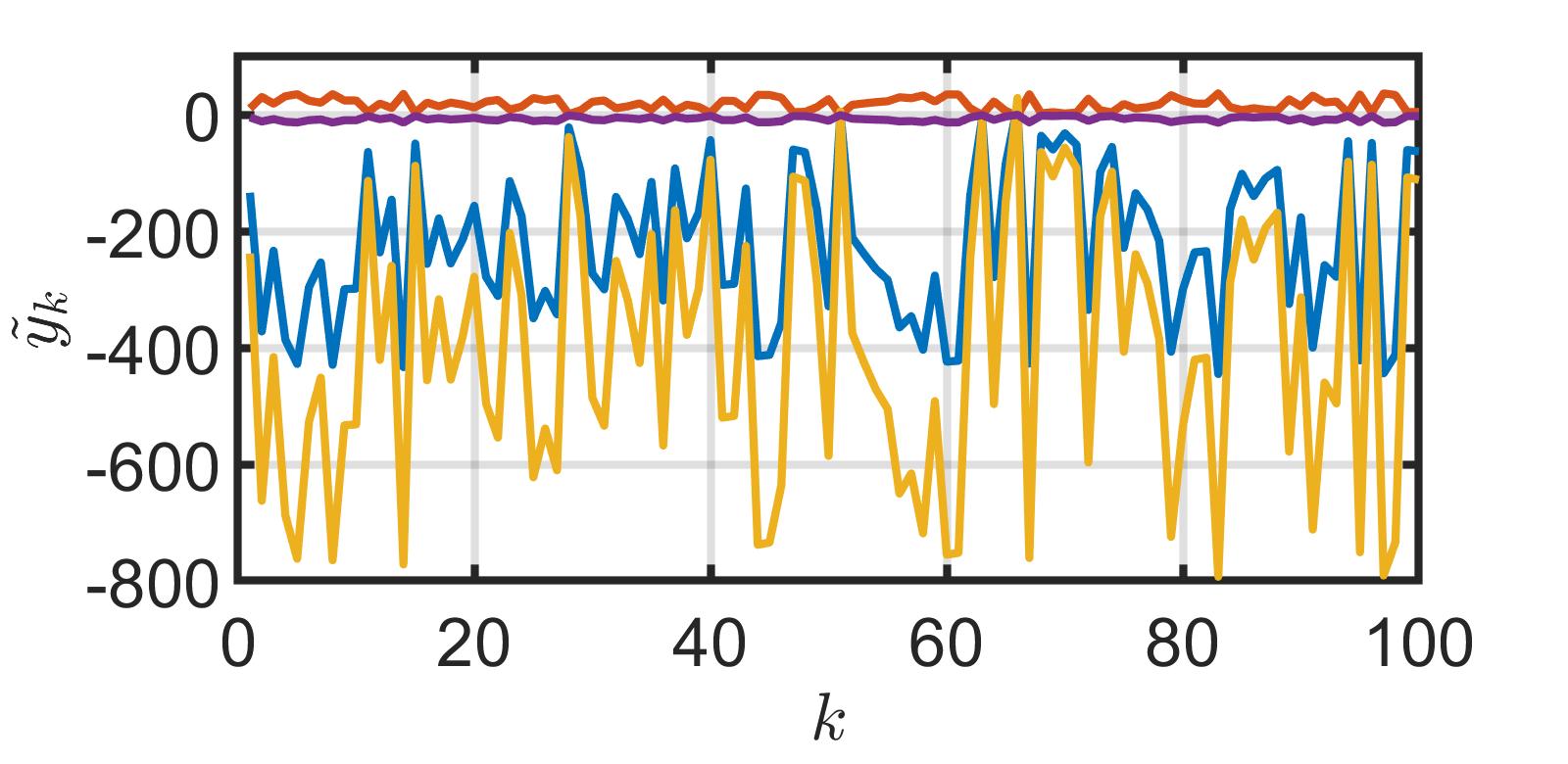}}
    \caption{Comparison between the measurement vector $y_k$ and the encoded measurements $\tilde{y}_k$.}
\label{yyp}
\end{figure*}
In Figure \ref{aap}, the original anomaly detector alarm $a_k$, the encoded alarm signal $\tilde{a}_k$, and the decoded alarm signal at the user $\hat{a}_k$ are compared. We assume that a deterministic additive fault $\delta=0.9$ is introduced into the system from $k=20$. As can be seen in this figure, the original alarm signal $a_k$ is a vector of zeros and ones, and it turns to $1$ from $k=20$, with some misdetections. Then, the encoded alarm $\tilde{a}_k$ is depicted, which is completely different from the original one and based on its random term $\Pi_9 b_k^1$. Besides, the dimension of the alarm signal is increased from one to two in the encoded alarm signal $\tilde{u}_k$. Then, the alarm is decoded at the user side as $\hat{a}_k$, which is the same as the original alarm ${a}_k$. Therefore, although encoded private signals $\tilde{y}_k$, $\tilde{u}_k$, and $\tilde{a}_k$ are totally different from original signals $y_k$, $u_k$, and $a_k$, the exact alarm signal can be encoded at the user side, and the proposed privacy-preserving anomaly detection algorithm has the same detection performance as standard anomaly detector. 
\begin{figure}[ht]
\begin{subfigure}{.5\textwidth}
  \includegraphics[width=3.3in]{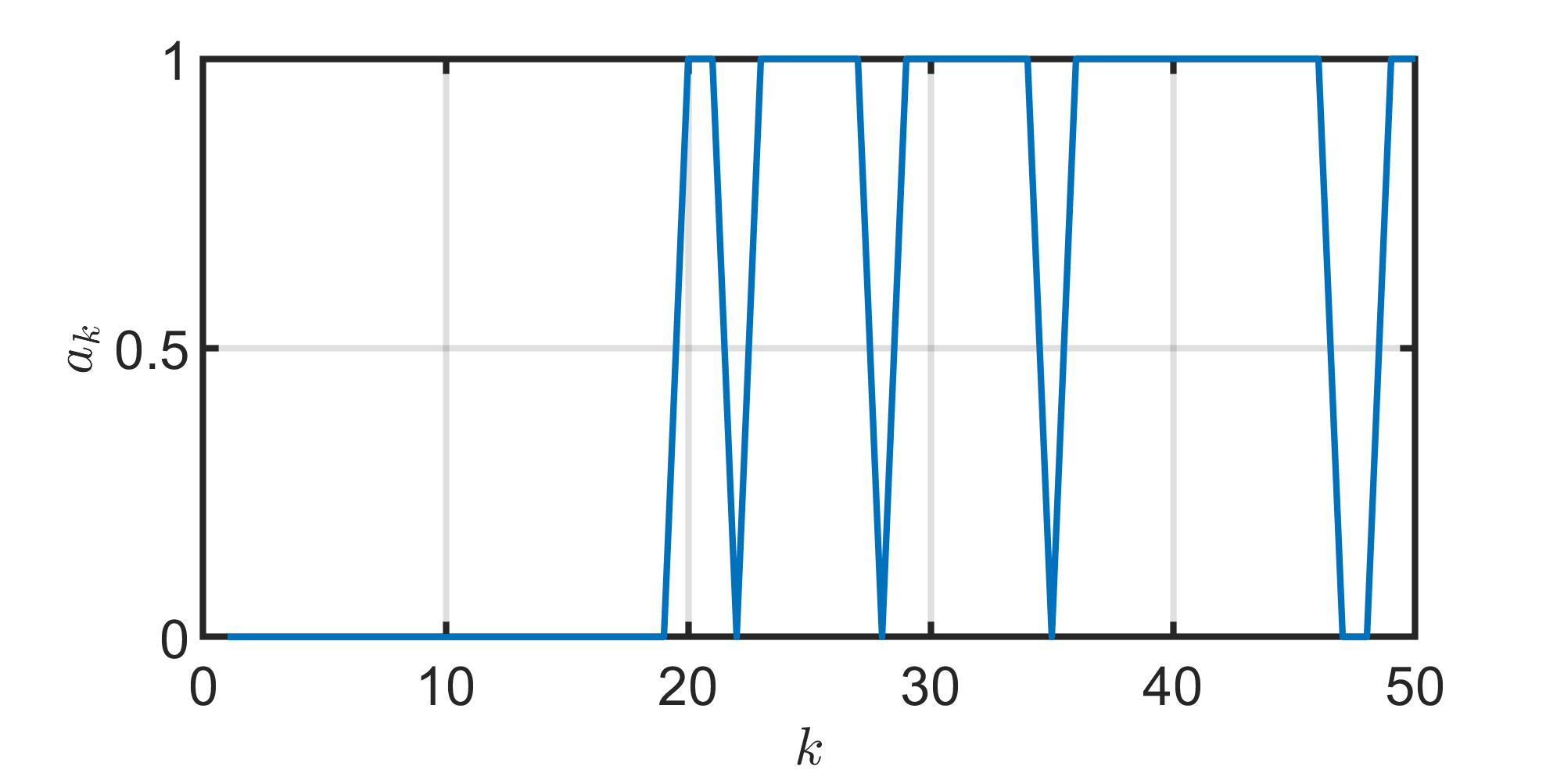}
\end{subfigure}
\begin{subfigure}{.5\textwidth}
  \includegraphics[width=3.3in]{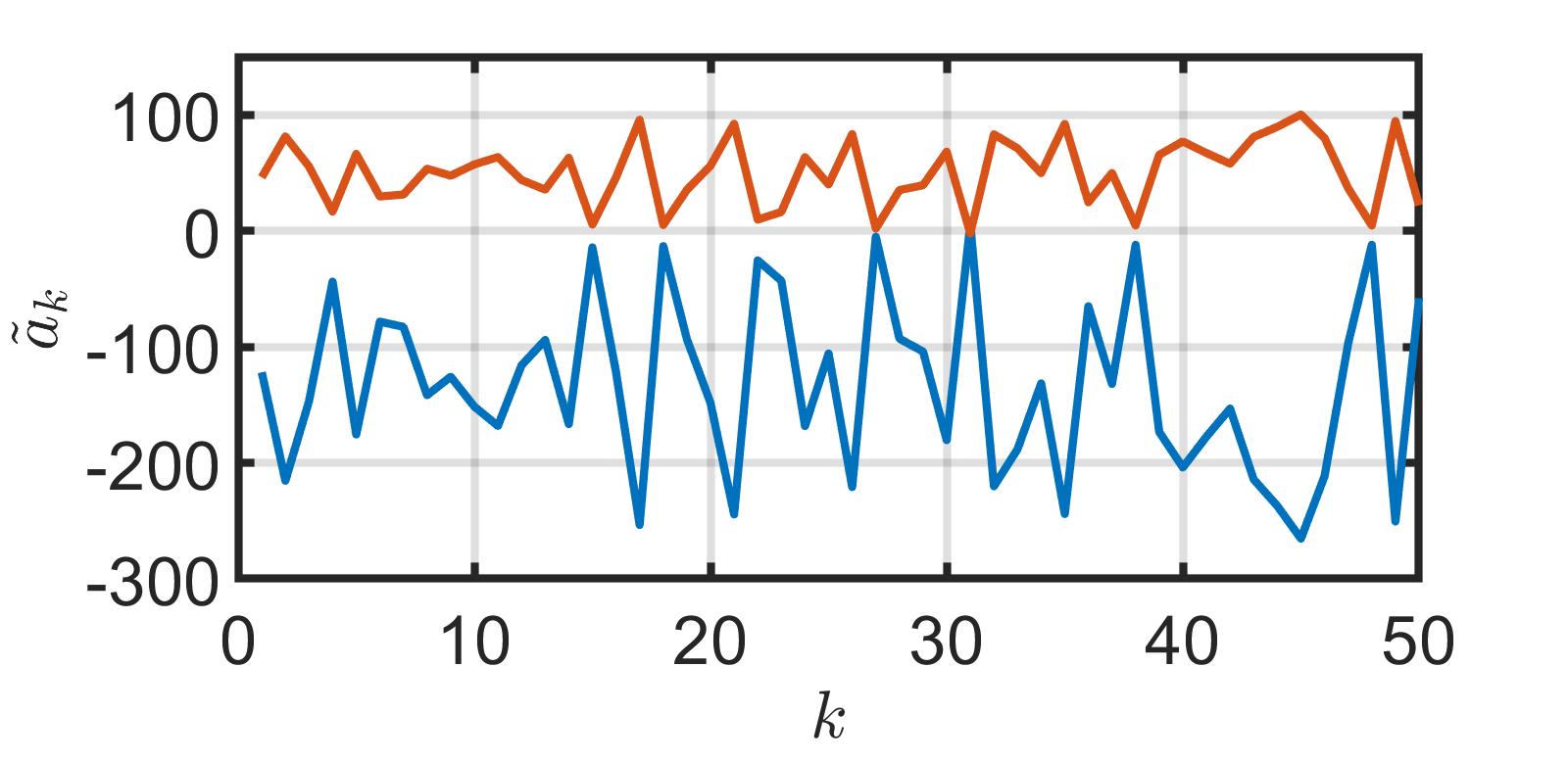}
\end{subfigure}
\begin{subfigure}{.5\textwidth}
  \includegraphics[width=3.3in]{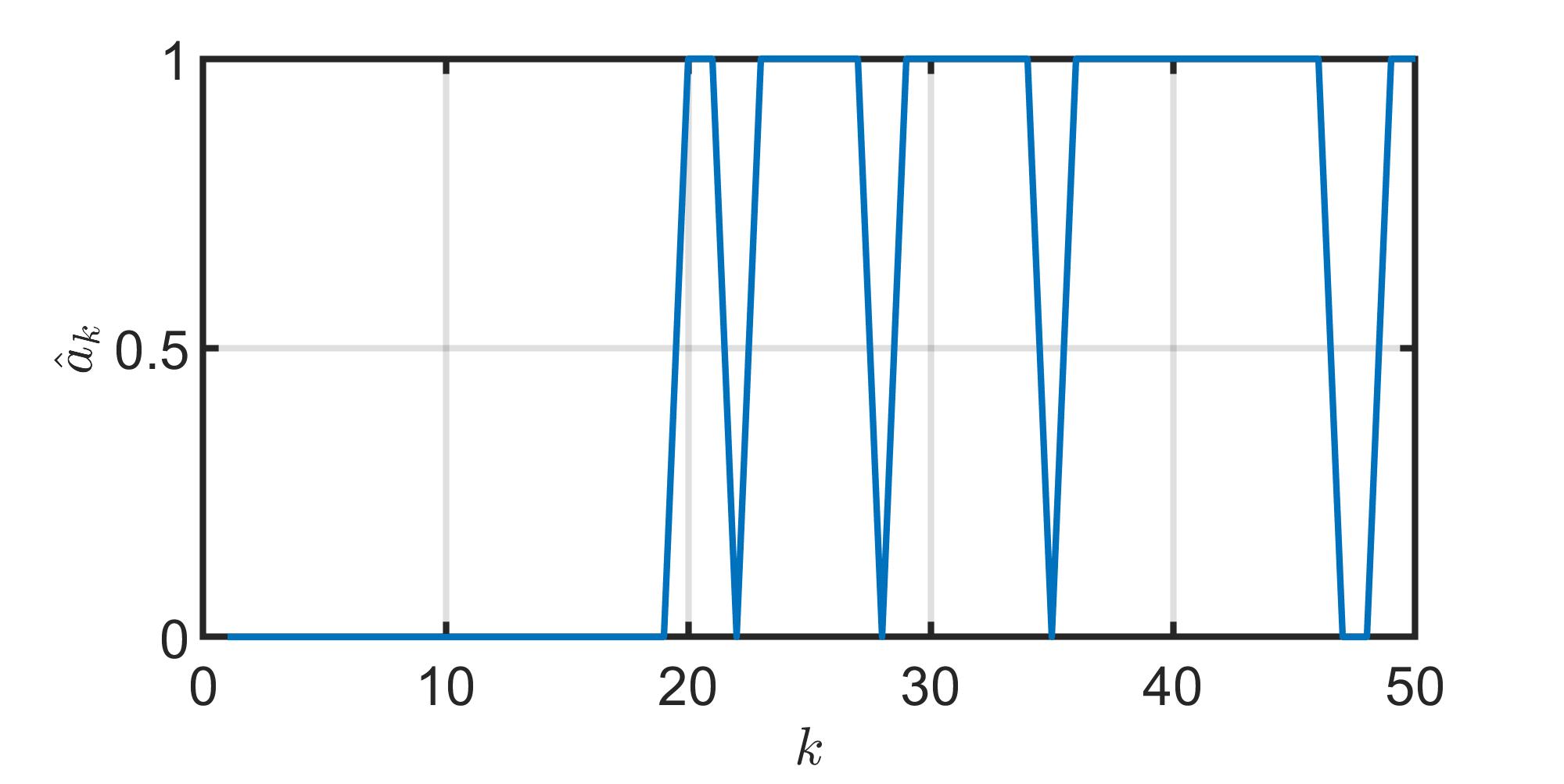}
\end{subfigure}
\caption{Comparison between the alarm signal $a_k$, the encoded alarm $\tilde{a}_k$, and the decoded one $\hat{a}_k$.}
\label{aap}
\end{figure}

\section{Conclusion}
In this paper, we have developed a privacy-preserving anomaly detection framework built on the synergy of random coding and immersion and invariance tools from control theory. We have devised a synthesis procedure to design the dynamics of a coding scheme for privacy and a target anomaly detector such that trajectories of the standard detector are immersed/embedded in its trajectories, and it works on encoded data. Random coding was formulated at the user side as a random change of coordinates that maps original private data to a higher-dimensional space. Such coding enforces that the target system produces an encoded version of the standard anomaly detector alarms.\\
The proposed (I\&I)-based anomaly detector provides the same performance as the standard anomaly detector, reveals no information about private data, and is computationally efficient. It provides a high level of privacy without degrading the anomaly detection performance. 
\bibliography{ifacconf32}
\end{document}